\documentclass{moriond}

\def\be{\begin{equation}}
\def\ee{\end{equation}}
\def\bea{\begin{eqnarray}}
\def\eea{\end{eqnarray}}

\newcommand{\Photo}{}

\begin{document}
\vspace*{4cm}
\title{COSMOLOGICAL CLUSTER TENSION}

\author{ A. BLANCHARD$^1$, Z. SAKR$^{1,2}$, S. ILI\'C$^{1,3}$ }

\address{$^1$IRAP, Universit\'e de Toulouse, CNRS, CNES, UPS, Toulouse, France.\\ $^2$ Universit\'e St Joseph; UR EGFEM, Faculty of Sciences, Beirut, Lebanon.\\
$^3$CEICO, Institute of Physics of the Czech Academy of Sciences, Na Slovance 2, Praha 8 Czech Republic}

\maketitle\abstracts{
The abundance of clusters is a classical cosmological probe  sensitive to both the geometrical aspects and the growth rate of structures. The abundance of clusters of galaxies measured by Planck has been found to be in tension with the prediction of the $\Lambda$CDM models normalized to Planck CMB fluctuations power spectra. The same tension appears with  X-ray cluster local abundance.  Massive neutrinos and modified gravity are two possible solutions to fix this tension. Alternatively,  others options include a bias in the selection procedure or in the  mass calibration of clusters.  We  present a study, based on our recent work\cite{sib}, updating the present situation on this topic and discuss the likelihood of the various options. }

\section{Introduction}

The $\Lambda$CDM scenario has become the standard scenario of modern cosmology, thanks to its quantitative agreement with several major observational results\cite{w13}, on top of which its good agreement with the angular power spectrum of the CMB fluctuations as measured by Planck\cite{P16cp}, and its ability to predict the large scale distribution of matter as measured by the correlation function of galaxies\cite{bdrs2}.\\
Despite of this success, some observables show to be in tension with the predictions of the  $\Lambda$CDM when normalized to the Planck CMB data. This is probably to be expected given the high accuracy of the modern observations relevant to cosmology, but it is impornat to figure out its possible origin. This may have to do with some
unidentified residual systematics in at least one set of observations or may be the signature for the need of a fundamental modification of the standard scenario, i.e. a hint for new physics. \\
The CMB fluctuations provide a direct estimation of the amplitude of matter fluctuations at $z \sim 1100$ which can be extrapolated down to redshift zero through the linear growth rate of the model. The tighest constraint obtained from Planck (including CMB lensing) is $\sigma_8 = 0.8150 \pm 0.0087$ (68\%). Several obsevations  from the low redshift universe are however indicative of a lower amplitude of matter fluctuations. \\

\begin{figure}
\centerline{\includegraphics[width=0.5\linewidth]{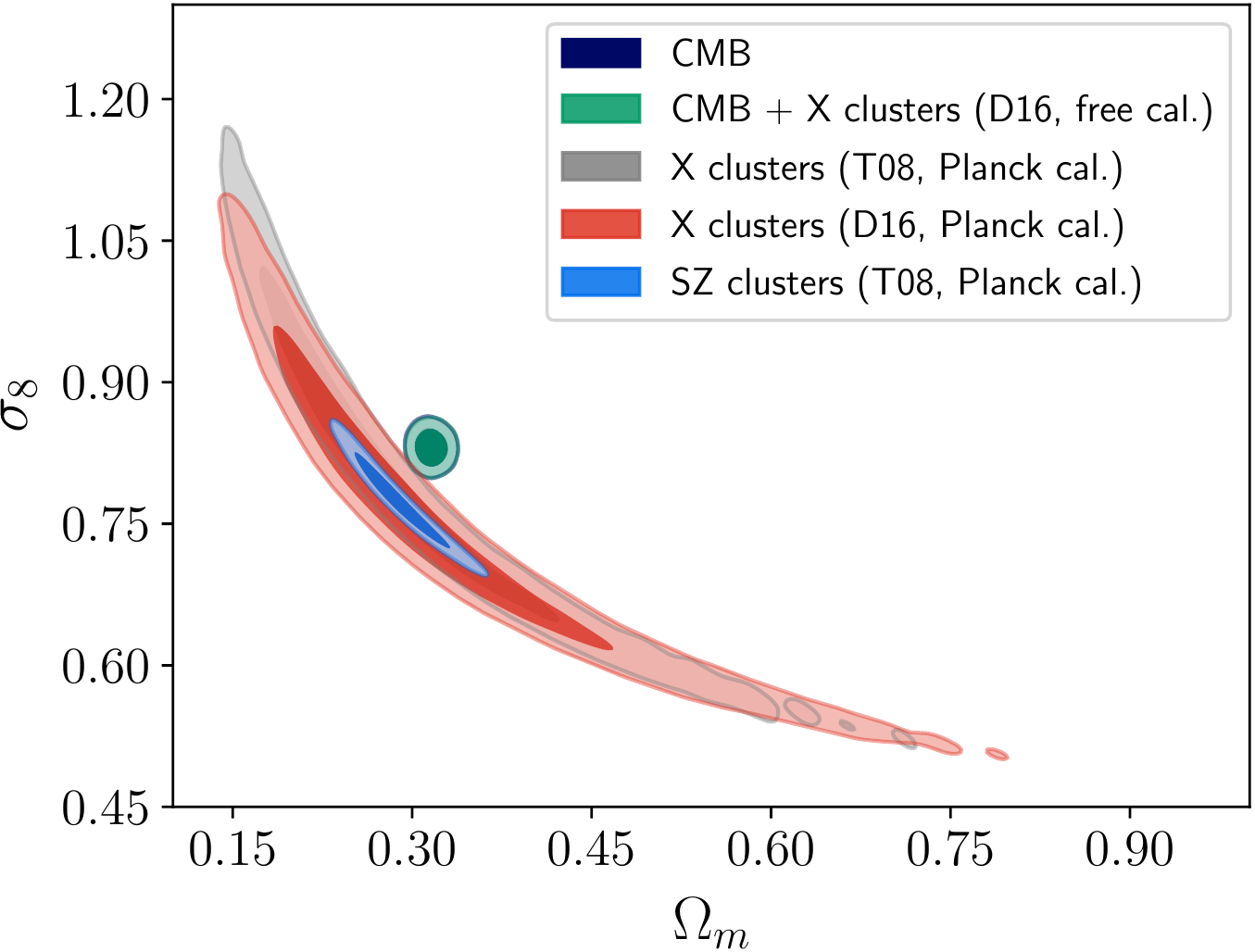}}
\caption[]{This figure illustrates the cluster tension: in the $\Omega_m-\sigma_8$ plane the contours obtained from the CMB alone (or with a free $A_{TM}$ calibration) do not overlap. The choice of the fitting function for the mass function (T08 or D16) has negligible effect. }
\label{fig:oms8}
\end{figure}

The abundance of clusters provides one way to constrain the amplitude of matter fluctuations. The number counts of clusters detected through their Sunyaev-Zel'dovich imprint as found by Planck lead to a lower amplitude $\sim 0.75$ for the same $\Omega_m$ with a specific value of the mass calibration. Although the abundance of local x-ray clusters provides a less stringent
tension, it yields an amplitude  $\sigma_8 \sim 0.75$ similar to the one derived from SZ counts, indicative that selection procedure is not an issue. This tension is illustrated in Fig. \ref{fig:oms8}. In the following, according to our recent work\cite{sib}, we provide details on the methodology on how clusters are modelled to establish this result. 

\section{How to use clusters for cosmology}
\subsection{The mass function}
Within a given cosmological model from a specific framework, it should be currently possible to compute the linear matter power spectrum at any epoch. This allows to compute the angular power spectrum of the CMB fluctuations as well as the large scale properties of the galaxies  distribution (provide one assumes a bias model). Knowing the power spectrum $P(k)$ one can compute the amplitude of matter fluctuations after a smoothing of the field by mean of a window function of scale $R$. The mass of structures on scale $R$ correspond to the mass enclosed  by the window function, which for a spherical top-hat window  is just $4/3 \rho_m \pi R^3$. Going from the linear amplitude of matter fluctuations $\sigma(m,z)$ is possible thanks to the magic of the (extended) Press and Schechter approach. From general arguments the non-linear mass function of objects resulting from gravitational collapse can be written in the form:

  \be
 n(M,z)  = - \frac{\overline{\rho}}{M^2\sigma(M)} \delta_{NL}(z) \frac{d\ln \sigma \;}{d\ln M}{\cal F}(\nu_{NL}) 
 \ee
i.e. a scaling law with mass and redshift\cite{bvm}. In the case of standard gaussian fluctuations, the mass function has been the subject of numerous numerical studies and analytical expressions for the function $\cal F$ have been proposed providing an accurate fit to the mass function inferred from CDM simulations. The situation has been slightly obscured by the fact that different definitions were used for the definition of an ``object'' and claims for departures from scaling of the mass function. However, theses differences are very minor in light of the above tension. The Tinker et al.\cite{Tinker} fit has been widely used. Despali et al.\cite{Despali} provided a new analysis showing  that standard scaling is preserved when virial radius is used while departures appear when different mass definitions are used (like the most used $M_{500}$). 

 \subsection{From mass to observable}

 In order to compare predictions from a specific model to observations one has to specify the relation between the mass and the observable. Such a relation can be deduced from scaling arguments\cite{Kaiser86}. When applied to gas temperature of x-ray clusters, this reads :
 \be T=A_{T-M}(h\, M_\Delta)^{2/3}\left(\frac{\Omega_{m} \Delta(z)}{178}\right)^{1/3}(1+z)
 \ee
$A_{T-M}$ being a calibration. The amplitude of the theoretical temperature distribution function is strongly sensitive to $\Omega_m$ and $\sigma_8$, strictly independent on $h$ and weakly dependent on the shape of the power spectrum $P(k)$ (it's also depends on the gaussian or not nature of the fluctuations).

\begin{figure}
\centerline{\includegraphics[width=0.5\linewidth]{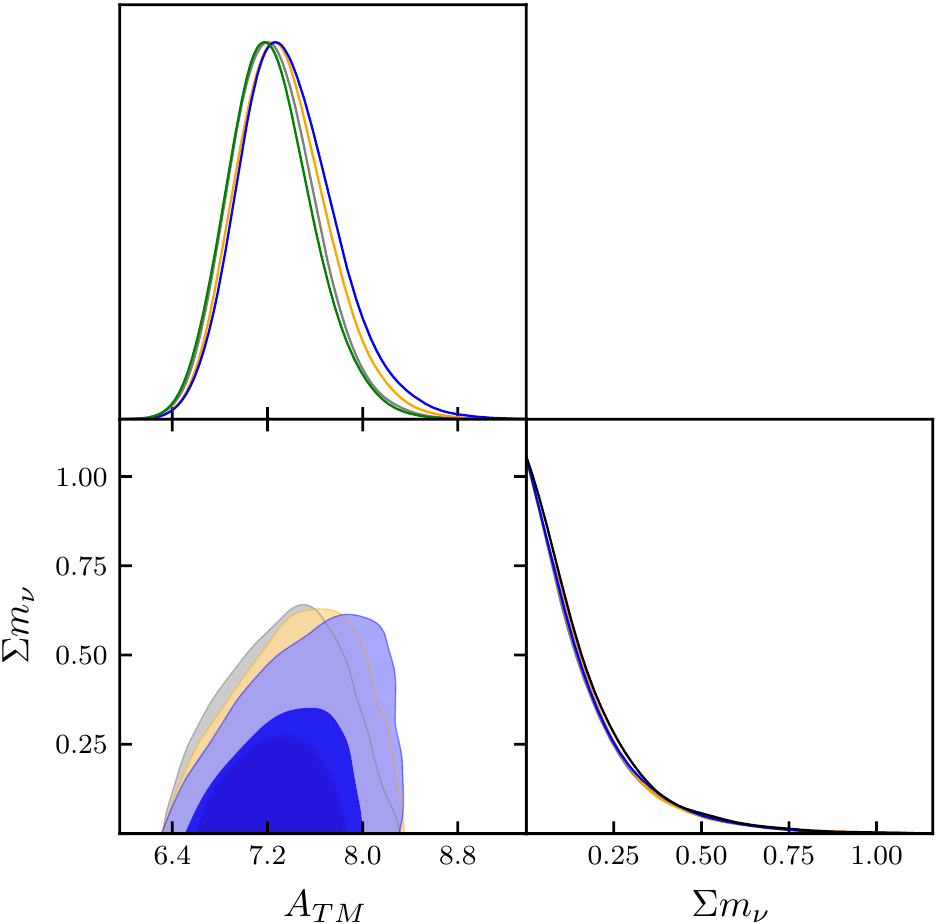}}
\caption[]{The calibration $A_{TM}$ does not appear to be correlated with a possible non zero neutrino mass when x-ray cluster constraint is combined with the CMB. Different prescription for the mass function in the presence of massive neutrinos does not lead to appreciable differences. Both liklihoods on $A_{TM}$
  and on neutrino mass are essentially unchanged compared to the massless case (and when using different prescriptions).  }
\label{fig:ATMmn}
\end{figure}

The abundance of local x-ray clusters can then be used as a powerful constraint on the parameters ($\Omega_m$, $\sigma_8$), but the relation is degenerated with the calibration. Once normalized to present day data, the redshift evolution essentially relies on the (linear) growth rate of fluctuations, making clusters abundance a non-geometrical powerful cosmological test. This test can be implemented from clusters detected by various technics. Applications to local x-ray clusters already showed some puzzling features\cite{bd05}. However, the most famous recent example is certainly the  cluster number counts obtained by Planck through the SZ effect. Indeed taken at face value the observed counts are lower  by a factor 3--4 than  expectations from the best  $\Lambda$CDM fitting CMB. However this tension relies entirely on the assumption, or prior, on the calibration: if the calibration is let free both SZ counts and local x-ray abundance can be fitted with the same calibration : $A_{T-M}\sim 7.3 \pm 0.3$ (at the virial radius), corresponding to $1-b \sim 0.6$ in Planck convention)\cite{ibd}, while Planck standard calibration is arround 8.7 corresponding to $1-b =  0.8$. 
A solution to solve this tension is to advocate a massive neutrino contribution that would alter the matter power spectrum, leading to a lower $\sigma_8$. We have examined in detail this possibility by running MCMC chains on CMB + local abundance of x-ray clusters with a free calibration. Our results showed that the likelihood on neutrino mass $m_\nu$  is unchanged
and that no correlation between  $A_{TM}$ and $m_\nu$  shows up in Fig.\ref{fig:ATMmn}. Identically the likelihood on $A_{TM}$ is unchanged.\\

\begin{figure}
\centerline{\includegraphics[width=0.45\linewidth]{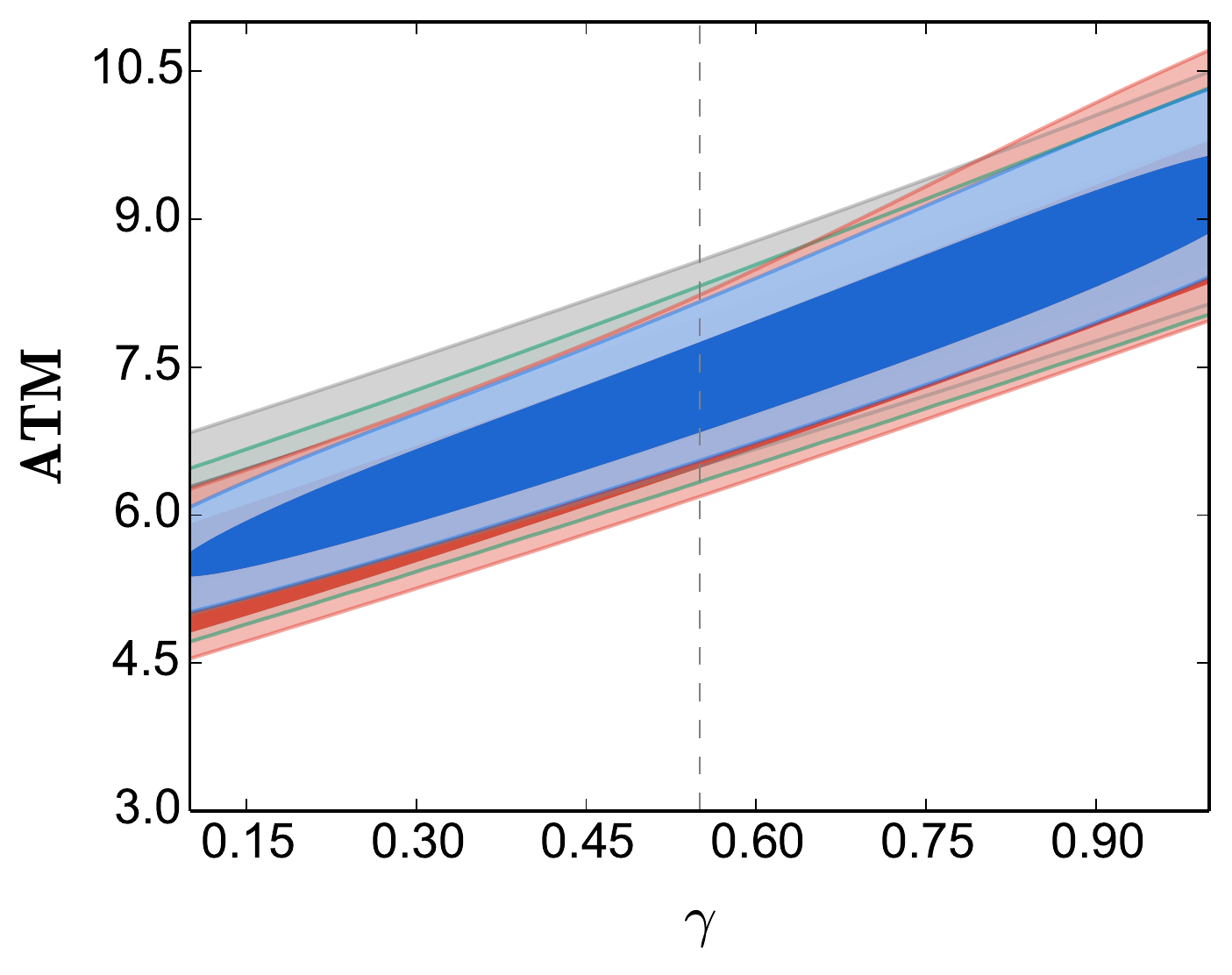}}
\caption[]{ The calibration $A_{TM}$  is tighly correlated to the parameter $\gamma$ in a simple representation of a modified theory of gravity. The correlation with massive neutrino (grey contour) is essentially the same to the massless case (green). The addition of the BAO and $Ly\alpha$, respectively red and blue, lead to very similar contours.}
\label{fig:ATMgamma}
\end{figure}

As an alternative we examine whether a modified gravity model, represented by a simple $\gamma$ growth rate, could solve the issue. Not surprisingly, we found that this possibility can indeed restore consistency between the Planck calibration and clusters counts in CMB normalized cosmology, but at the expense of a large value of  $\gamma$ (of the order to $0.9\pm 0.1$), with a tight correlation between $A_{TM}$  and  $m_\nu$, independent of the details of the models or additional constraints used, see Fig.\ref{fig:ATMgamma}. 

\section{Conclusion}

The CMB-cluster tension, consistently appearing in SZ and x-ray, relies uniquely on the cluster mass calibration used in scaling laws. We found that massive neutrinos does not alleviate the tension while a modified gravity model  represented by a $\gamma$ parametrization of the growth rate can accomodate both data sets provide $\gamma \sim 0.9\pm 0.1$. We conclude that if the standard Planck calibration $1-b \sim 0.8$ is reliabily confirmed, it would provide a strong indication of some form of exotic physics in the dark sector.

\section*{Acknowledgments}

 We acknowledge C.~Y\`eche and collaborators for providing us their code to implement \mbox{Lyman-$\alpha$} constraints.

\end{document}